# Adaptive Control of an Inverted Pendulum by a Reinforcement Learning-based LQR Method

Uğur Yıldıran

Yildiz Technical University, Control and Automation Engineering Department,
İstanbul, Turkey
uyildiran@yildiz.edu.tr

**Abstract**

Inverted pendulums constitute one of the popular systems for benchmarking control algorithms. Several methods have been proposed for the control of this system, the majority of which rely on the availability of a mathematical model. However, deriving a mathematical model using physical parameters or system identification techniques requires manual effort. Moreover, the designed controllers may perform poorly if system parameters change. To mitigate these problems, recently, some studies used Reinforcement Learning (RL) based approaches for the control of inverted pendulum systems. Unfortunately, these methods suffer from slow convergence and local minimum problems. Moreover, they may require hyperparameter tuning which complicates the design process significantly. To alleviate these problems, the present study proposes an LQR-based RL method for adaptive balancing control of an inverted pendulum. As shown by numerical experiments, the algorithm stabilizes the system very fast without requiring a mathematical model or extensive hyperparameter tuning. In addition, it can adapt to parametric changes online.

## 1. INTRODUCTION

An inverted pendulum is an underactuated system for which the goal is to stabilize a rod around the unstable equilibrium at the upright position. There are different variants of it such as single pendulum, double pendulum, the pendulum on a chart, and rotary pendulum [1]. This system constitutes one of the important benchmarks for control algorithms due to its instability and nonlinearity. Moreover, it is representative of some important real-life problems including human walking, rocket guidance, and balancing scooters.

Inverted pendulum systems have been extensively studied in the literature, and various control methods have been implemented to stabilize them. Linear output or state feedback methods, such as PID or LQR control, were applied in [2-4] while sliding mode control was used in [5] for robust stabilization. A fuzzy control algorithm was employed in [6], and a nonlinear Model Predictive Control approach was developed in [7]. Recently, an Active Disturbance Rejection strategy was proposed in [8]. Hybrid algorithms combining different methods were also explored in [1, 4, 9, 10]. Although these studies achieved stabilization and satisfactory performance, they rely on a mathematical model of the system. Such models can be obtained by applying first principles or using system identification techniques. However, modeling is a time-consuming task requiring human effort. Furthermore, derived models may not be valid if there are changes in the system over time, leading to performance degradations or instabilities.

Motivated by these complications, some researchers employed Reinforcement Learning (RL) techniques for the control of inverted pendulum systems. In [11], a batch reinforcement learning method was proposed for a wheeled pendulum robot. The underlying Q-learning algorithm is based on a finite Markov Decision Process (MDP) framework, which requires discretization of state and action spaces and represents the Q-function in a tabular form. The paper [12] compared different RL algorithms applied to an inverted pendulum. Similarly, they approximate the continuous system as a finite MDP. Due to discretizations, the methods investigated in [11, 12] suffer from the well-known curse of dimensionality problem.

In the last years, another line of research tried to benefit from function approximations to alleviate the curse of dimensionality problem [13–16]. These papers utilized Deep Neural Networks (DNN) for representing actors and critics. Parameters of DNNs were updated through policy gradient algorithms to find the best Q-function approximation and policy corresponding to system dynamics and reward function. Although pendulums could be stabilized, training was too slow and took many episodes to converge. Moreover, it may be necessary to make many trials to set hyperparameters properly and get rid of local minima. Online adaptation also seems to be problematic due to these reasons.

LQR is a well-known method for the optimal control of dynamic systems. The corresponding policy has a simple linear form, and the associated value function (also the Q-function) can be shown to be quadratic. Thus, for the LQR problem, the optimal actor and critic have simple forms [17]. This alleviates the need for using complex function approximations. Consequently, one can expect significant speed-ups in the training process. Moreover, hyperparameters of DNNs and their tuning can be eliminated. With this observation in mind, in [18], a simple and efficient LQR-based Q-learning algorithm was proposed. This approach gained significant interest very recently [19–21].

The LQR method was demonstrated to be successful in stabilization of inverted pendulum systems in past studies as mentioned above. Moreover, it is possible to devise an RL counterpart of this method for fast leaning-based control as discussed. Motivated by these facts, in the present study, an LQR-based RL algorithm is developed and implemented for optimal adaptive control of an inverted pendulum system. The algorithm is elaborated and its success is demonstrated by simulations.

The paper is organized as follows. The inverted pendulum model is introduced in Section 2. The proposed LQR-based RL algorithm is described in Section 3. Simulation results verifying its stability, convergence, and adaptation capabilities are presented in Section 4. The main findings are discussed in Section 5.

## 2. MATHEMATICAL MODEL

The inverted pendulum system considered in this study is depicted in Figure 1. As can be seen from the figure, the system is composed of a chart and a pendulum attached to it. The mass of the pendulum, $m$, is represented as a point mass located at the end of the rod. Chart mass, chart position, and pendulum angle are denoted as $M, y$, and $\theta$, respectively. The force input is shown as $u$. Friction forces are neglected.

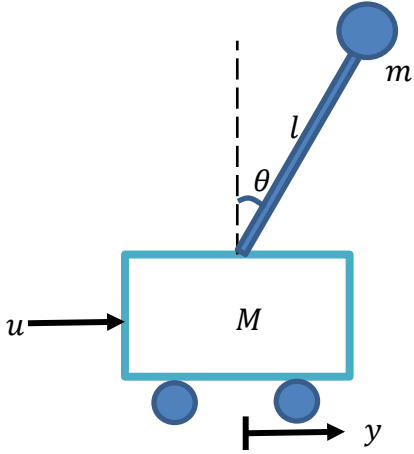

Figure 1 Inverted pendulum system

The mathematical model of the system can be obtained by deriving associated Lagrangian equations. Since this procedure is well known, the details are skipped, and the final model is given below. The reader is referred to [3] for derivations.

$$\ddot{\theta} = \frac{u\cos\theta - (M+m)g\sin\theta + ml\dot{\theta}^2\cos\theta\sin\theta}{ml\cos^2\theta - (M+m)l}$$

$$\ddot{y} = \frac{u + ml\dot{\theta}^2\sin\theta - mg\sin\theta\cos\theta}{M+m-m\cos^2\theta}$$

In the above, $g$ represents the gravitational constant. By defining the state vector as $x = [\theta, \dot{\theta}, y, \dot{y}]$, the state space model can be expressed as

$$\begin{aligned}
\dot{x}_1 &= x_2 \\
\dot{x}_2 &= \frac{u\cos x_1 - (M+m)g\sin x_1}{ml\cos^2 x_1 - (M+m)l} \\
&\quad + \frac{ml\, x_2^2 \cos x_1 \sin x_1}{ml\cos^2 x_1 - (M+m)l} \\
\dot{x}_3 &= x_4 \\
\dot{x}_4 &= \frac{u + mlx_2^2\sin x_1 - mg\sin x_1 \cos x_1}{M+m-m\cos^2 x_1}
\end{aligned}$$

In this work, a discrete-time approach will be employed for the control of the inverted pendulum based on Q-learning. Thus, the state space model introduced above will be discretized for controller implementation. The Euler approximation method will be used for this purpose. The resulting system will have the following form

$$x_{t+1} = f_N(x_t, u_t), \qquad (1)$$

where $x_t \in \mathbb{R}^n$ and $u_t \in \mathbb{R}^m$. For the inverted pendulum system considered, one has $n = 4$ and $m = 1$.

## 3. PROBLEM FORMULATION

The goal is to stabilize the system around the unstable equilibrium point at the upright position without performing a swing up. This stabilization region can be expressed as

$$\mathcal{S} = \{x_t \in R^n | \underline{X} \le x \le \overline{X}\},$$

where $\underline{X}$ and $\overline{X}$ are vectors of lower and upper bounds in the state space such that the origin is contained in $\mathcal{S}$.

Since the RL algorithm will work in the vicinity of the equilibrium point $x_t = 0$, the system can be represented well by the following discrete-time linear dynamics, which corresponds to the linearization of the nonlinear dynamics given in (1).

$$x_{t+1} = f(x_t, u_t) = Ax_t + Bu_t$$

The stabilization problem can be formulated as a **deterministic** Markov Decision Process (MDP) with **continuous** state and action spaces. To be more specific, it can be represented by the tuple $(\mathcal{S}, \mathcal{A}, f, r, \gamma)$. Here, $\mathcal{S}$ is the state set defined above, $\mathcal{A} = \mathbb{R}$ is the action set, $f$ is the linear state equation introduced above, $r(x_t, u_t) = x_t'Qx_t + u_t'Ru_t$ is the quadratic reward function, and $\gamma$ is the discount factor.

The associated reinforcement learning task is to find deterministic policy $\pi: \mathcal{S} \to \mathcal{A}$ optimizing the following problem.

$$\min_{\pi} \sum_{t=0}^{\infty} \gamma^t r(x_t, u_t) \quad (2)$$
$$\text{s.t. } x_{t+1} = f(x_t, u_t)$$
$$u_t = \pi(x_t)$$

For the linear state equations and quadratic cost given above, and $\gamma = 1$, this becomes equivalent to the following Linear Quadratic Regulator (LQG) problem from control theory [22].

$$\min_{K} \sum_{t=0}^{\infty} x_t'Qx_t + u_t'Ru_t \quad (3)$$
$$\text{s.t. } x_{t+1} = Ax_t + Bu_t$$
$$u_t = Kx_t$$

Note that in the RL literature, the discount factor $\gamma$ is usually chosen smaller than one in order to ensure having a finite objective value for an optimal solution. However, it is a very well-known fact that the optimal solution of (3) is finite for a system having controllable dynamics [17]. Thus, we chose safely $\gamma = 1$, which corresponds to the formulation commonly accepted in the control literature.

The proposed algorithm will automatically learn the optimal state feedback gain corresponding to this problem by interacting with the system without making use of a mathematical model (i.e. system matrices $A$ and $B$ will not be available). Learning will comprise episodes that will be repeated till achieving stabilization. Each episode terminates when the pendulum states get out of $\mathcal{S}$.

## 4. LQR-BASED Q-LEARNING ALGORITHM

The proposed RL strategy is based on a Q-learning method. In the sequel, firstly Q-learning will be described. Then, its adaptation to LQR control will be introduced.

### 4.1. Q-Learning Method

Define the optimal infinite horizon value function associated with (2).

$$V(x_t) := \min_{\pi} \sum_{\tau=t}^{\infty} \gamma^\tau r(x_\tau, u_\tau)$$
$$\text{s.t. } x_{\tau+1} = f(x_\tau, u_\tau) \quad (4)$$
$$u_\tau = \pi(x_\tau)$$

The associated optimal Q-function (action-value function) that gives the minimum total reward after taking action $u_t$ can be defined as

$$Q(x_t, u_t) := r(x_t, u_t) + \gamma V(x_{t+1}), \quad (5)$$

where $x_{t+1} = Ax_t + Bu_t$.

The well-known Q-learning algorithm allows learning an optimal Q-function by interacting with the environment using the following update rule [17, 23].

$$Q(x_t, u_t) \leftarrow (1 - \alpha)Q(x_t, u_t) +$$
$$\alpha \left( r(x_t, u_t) + \gamma \min_{u_{t+1}}[Q(x_{t+1}, u_{t+1})] \right), \quad (6)$$

where $\alpha$ is the learning rate, which should satisfy $0 \leq \alpha \leq 1$. This rule updates the Q-

function $Q$ by taking a weighted average of its old value with the new target value appearing in the second term on the right-hand side (the term within the parentheses which is multiplied by $\alpha$). In this way, it can calculate expectations for stochastic problems statistically by performing filtering (temporal difference method). But for deterministic problems, like the LQR problem considered in this study, the learning rate can be taken as $\alpha = 1$ since there is no expectation.

If the optimal Q-function is known, the desired optimal policy can be obtained by solving the following optimization problem

$$\pi(x_t) = \underset{u_t}{\operatorname{argmin}} Q(x_t, u_t). \quad (7)$$

### 4.2. Q-learning for the LQR problem

As described in Section 3, for the LQR problem given in (3), the discount factor and learning rate can be taken as $\gamma = 1$ and $\alpha = 1$. Thus, the following simplified learning rule can be obtained from (6).

$$Q(x_t, u_t) \leftarrow r(x_t, u_t) + \underset{u_{t+1}}{\min}[Q(x_{t+1}, u_{t+1})] \quad (8)$$

To be able to apply this update rule, one needs to choose a proper representation for the Q-function. The simplest choice can be a tabular representation. It can be employed to approximate the Q-function by discretizing state and action spaces. But tabular representation makes the Q-learning algorithm impractical for high dimensional systems due to the curse of dimensionality problem. The space and time requirements grow exponentially with the number of dimensions.

Fortunately, for the LQR problem formulated in (3), it is a well-known fact that the Q-function can be expressed exactly as a quadratic function of state and action vectors without making any approximation [18]. In other words, it can be written in the following parametric form.

$$Q(x_t, u_t) = \begin{bmatrix} x_t \\ u_t \end{bmatrix}' M \begin{bmatrix} x_t \\ u_t \end{bmatrix}$$

where, $M \in R^{(n+m) \times (n+m)}$ is the symmetric parameter matrix. Since $M$ is symmetric, it has $(n + m + 1) \times (n + m)/2$ free parameters. This is very small when compared with the memory requirements of a tabular representation, which can represent the Q-function only approximately.

For the LQR problem, given the Q-function, one needs to obtain the corresponding policy by solving (7). This can be done conveniently using linear algebra techniques because the function $Q$ is a quadratic function of the action $u_t$. To this end, partition the parameter matrix $M$ as follows.

$$M = \begin{bmatrix} M_{11} & M_{12} \\ M_{21} & M_{22} \end{bmatrix},$$

where $M_{11} \in \mathbb{R}^{n \times n}$, $(M_{12})' = M_{21} \in \mathbb{R}^{m \times n}$, and $M_{22} \in \mathbb{R}^{m \times m}$. Then, it can be easily inferred that the $u_t$ minimizing (7) is attained at $u_t = Kx_t$, where $K = M_{22}^{-1} M_{21}$. Consequently, the learning rule (8) can be written as

$$Q(x_t, u_t) \leftarrow r(x_t, u_t) + \begin{bmatrix} x_{t+1} \\ Kx_{t+1} \end{bmatrix}' M \begin{bmatrix} x_{t+1} \\ Kx_{t+1} \end{bmatrix} \quad (9)$$

This identity will be employed to update the parameter matrix $M$ in the RL algorithm. To be more specific, the right-hand side of the equation will generate target values for the function $Q$ based on state observations and rewards received from experiments. One option is to employ each target value generated immediately to update the matrix $M$ using a gradient descent algorithm. But this method will bring an additional meta parameter, step-size, for which a proper value should be determined. In addition, gradient descent algorithm can hurt the stability of the overall system.

To overcome these complications, in the present study, a batch learning type approach is employed. As the system interacts with the environment, $n_s$ samples will be generated from observations for states and inputs in addition to target values generated by the right-hand side of (9), which will be denoted as follows.

State samples: $x_{s:s+n_s-1} =: (x_s, \ldots, x_{s+n_s-1})$,

Input samples: $u_{s:s+n_s-1} =: (u_s, \ldots, u_{s+n_s-1})$,

Target samples:
$$q^{tar}_{s:s+n_s-1} := (q^{tar}_s, \ldots, q^{tar}_{s+n_s-1}),$$

where $s$ is the start of the sampling window.

Then, these samples are used to construct the following set of equations whose solution gives the parameter matrix $M$ of the updated Q-function appearing on the left-hand side of (9).

$$\begin{bmatrix} x_\tau \\ u_\tau \end{bmatrix}' M \begin{bmatrix} x_\tau \\ u_\tau \end{bmatrix} = q^{tar}_\tau, \tau = s:s + n_s - 1$$

Using matrix algebra, these equations can be expressed as

$$\text{vec}(M)\left(\begin{bmatrix} x_\tau \\ u_\tau \end{bmatrix} \otimes \begin{bmatrix} x_\tau \\ u_\tau \end{bmatrix}\right) = q^{tar}_\tau, \quad (10)$$
$$\tau = s:s + n_s - 1,$$

where $\otimes$ represents the Kronecker product operator and $\text{vec}(M)$ is the row vector obtained by stacking the rows of matrix $M$ horizontally.

Clearly, (10) is a set of linear equations in elements of $M$. It is known that under an $\epsilon$-greedy exploration strategy with large enough $\epsilon$, they will be linearly independent [24]. Thus, one can find a unique solution by choosing $n_s \geq (n + m + 1) \times (n + m)/2$ because the number of equations will be at least as much as the number of unknowns. The matrix $M$ can be found by solving the following least squares optimization problem.

$$\min \quad \frac{1}{2} e'e$$
$$\text{s.t.} \quad A \, \text{vec}(M)' - b = e,$$

where $A$ is the matrix whose rows are obtained by stacking the row vectors $[x'_\tau, u'_\tau] \otimes [x'_\tau, u'_\tau], \tau = s:s + n_s - 1$, and $b$ is the column vector whose elements are $q^{tar}_\tau, \tau = s:s + n_s - 1$. The solution is given by the following equation

$$\text{vec}(M)' = (A'A)^{-1}A'b. \quad (11)$$

### 4.3. Proposed Algorithm

Making use of material presented in Section 4.2, one can obtain the algorithm given in Figure 2.

The algorithm starts by initializing state vector $x_0$, matrix $M$, time index $t$ and sample window start time index $s$. The initial control gain is also computed in line 4 from the initial $M$ matrix.

This is followed by the while loop which is executed throughout the experiment. The loop is composed of three blocks.

In the first block, one time step of the experiment is executed as follows. In line 6, the gain $K$ is multiplied by the state vector $x_t$ and a random exploration noise $\varepsilon$ is added to compute the input $u_t$ from the $\epsilon$-greedy policy. Then, reward and next state are calculated in lines 7 and 8 from the applied input $u_t$ and state observation $x_t$. These are used in line 9 to obtain a target value for the $Q$ function.

The second block of the loop is for updating the Q-function. More specifically, after every $n_s$

iteration, the algorithm executes the body of the if statement. In this part, the new $M$ matrix is computed by solving (11) making use of input, state, and reward observation collected in the last $n_s$ time steps.

```
1  M = init_M()
2  x_0 = init_x()
3  t, s = 0
4  K = -M_22^{-1} M_21
5  while True do
      // One step simulation
6     u_t = K x_t + ε
7     r_t = x_t' Q x_t + u_t' R u_t
8     x_{t+1} = f(x_t, u_t)
9     q_t^{tar} = r_t + [x_{t+1}; K x_{t+1}]' M [x_{t+1}; K x_{t+1}]
      // Q-function update
10    if t == s + n_s - 1 then
11       M ← solution of (11)
12       K = -M_22^{-1} M_21
13       s = t + 1

      // Episode termination
14    if not X_ ≤ x ≤ X̄ then
15       x = init_x()
16    if ||M|| > H then
17       M = init_M()
18       x = init_x()
19    t = t + 1
```

Figure 2 Proposed LQR-based Q-learning algorithm

The third part comprises two termination criteria. In the first one, if the state vector gets out of the state set $\mathcal{S}$ defined by the lower bound $\underline{X}$ and the upper bound $\overline{X}$, it is reset to an initial position. Similarly, the second criterion checks whether the matrix $M$ diverges. If the norm of $M$ gets larger than a chosen threshold $H$, it is reset to an initial matrix.

There are two functions, namely init_x and init_M, used in the algorithm to reset states and the parameter matrix $M$. Their pseudocodes are given below.

```
function init_x():
    x_0 = [1, 0, 0, 0] * rand() * ν
    return x_0
function init_M():
    M_0 = μ [Q 0; 0 R]
    return M_0
```

init_x function returns a state whose value is close to the upright position. Here, rand() is a function that generates a uniformly distributed random number in the interval [-0.5,0.5] while $ν$ is a small constant value. This procedure represents a manual initialization of the pendulum by the operator to the upright position, which cannot be performed perfectly, resulting in deviations from the ideal state.

init_M function returns a block diagonal matrix whose diagonal elements are $Q$ and $R$. This matrix is multiplied by a scaling constant μ. This choice is observed to work well in general for several experiments.

## 5. SIMULATION RESULTS

The algorithm introduced in the previous section was applied to the nonlinear inverted pendulum system described in Section 2. The model parameters were chosen as $m = 0.2$ kg, $M = 0.5$ kg, $l = 0.3$ m, and $g = 9.8$ m/s². Quadratic cost matrices were chosen as

$$Q = \begin{bmatrix} 100 & 0 & 0 & 0 \\ 0 & 1 & 0 & 0 \\ 0 & 0 & 10 & 0 \\ 0 & 0 & 0 & 1 \end{bmatrix}, \quad R = 1.$$

The scaling factor used for initializing the matrix $M$ was taken as $\mu = 10$, which was observed to work well in general. The constant used in init_x function was chosen as $\nu = 5 \times 10^{-3}$.

Two experiments were performed. In both, the proposed algorithm was applied to the nonlinear pendulum model, not to its linearization, for learning optimal controller gains stabilizing the system. To show how close the computed controllers are to ideal ones, the corresponding LQR gains were also calculated making use of the system matrices $A$ and $B$ which were obtained by linearizing the model. The results are elaborated below.

In the first experiment, the Q-learning algorithm was run to learn controller parameters from scratch. The norm of the difference between the controller gain computed by the algorithm and the optimal gain obtained by linearization is shown in Figure 3. In addition, time evaluations of states and the input are depicted in Figure 4.

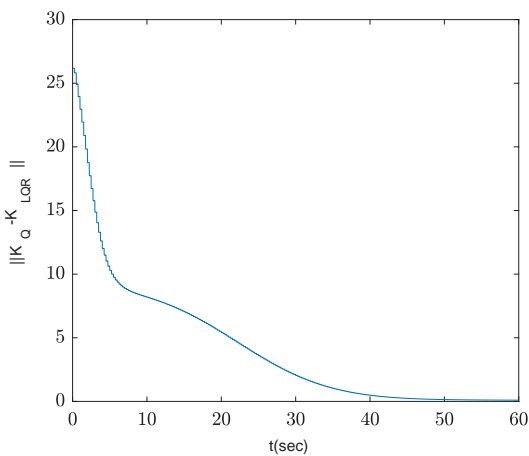

Figure 3 The norm of the difference between feedback gains computed by the Q-learning ($K_Q$) and model-based ($K_{LQR}$) LQR methods

As can be seen from Figure 3, controller gains computed online by the Q-learning algorithm converged to optimal LQR gains, which were calculated by making use of the linearized model, after 50 seconds. This is achieved in two epochs, the second of which starts at t=4.73 seconds after peaks appearing in Figure 4. These peaks occurred since the algorithm diverged in the first epoch after which the system is reset to start the second epoch. States converge to the desired value within 30 seconds starting from the beginning of the experiment and excluding the time for bringing the pendulum to the initial position after divergence, which should be performed manually in a real test bed. Note that the actual settling time of the optimal controller learned is much shorter than 30 seconds, and in fact, the same as that of the optimal model-based LQR controller because their gains are practically the same. (the norm of their difference converged to zero as mentioned above). These gains are found to be

$$K = [23.2855, 3.7400, 0.9185, 1.9712]$$

To demonstrate the adaptation capabilities of the proposed algorithm, a second experiment was conducted. Starting with the optimal controller gains found by the algorithm at t=0 seconds, a step change was applied to the model parameters. Specifically, at t=20 seconds, the pendulum was assumed to have broken by being cut in half, which was reflected in the model by halving the length and mass of the pendulum. As before, time evaluations of state variables and the distance of learned gains from optimal ones computed by the model-based LQR method are given in Figure 5 and Figure 6, respectively.

Figures show that controller gains and states converge rapidly (in around 10 seconds). This shows that the algorithm can adapt very quickly in response to even large parametric changes. Although controller gains initially exhibited

large deviation as can be seen from Figure 6, states are affected to a small extent as can be observed from Figure 5. This can be attributed to well-known robustness properties of LQR-based controllers. Controller gains after convergence are found as

$$K = [21.2353, 2.4408, 2.7611, 3.0821]$$

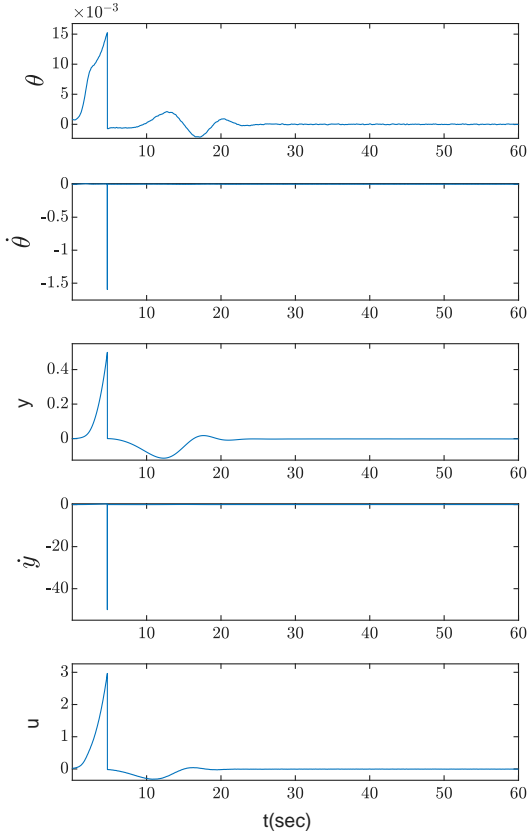

Figure 4 Time evaluations of states and control inputs for the Q-learning-based LQR method

computationally intensive processes necessary for updating DNN parameters. This computational burden is compounded by the fact that multiple experiment repetitions are often necessary to tune hyperparameters.

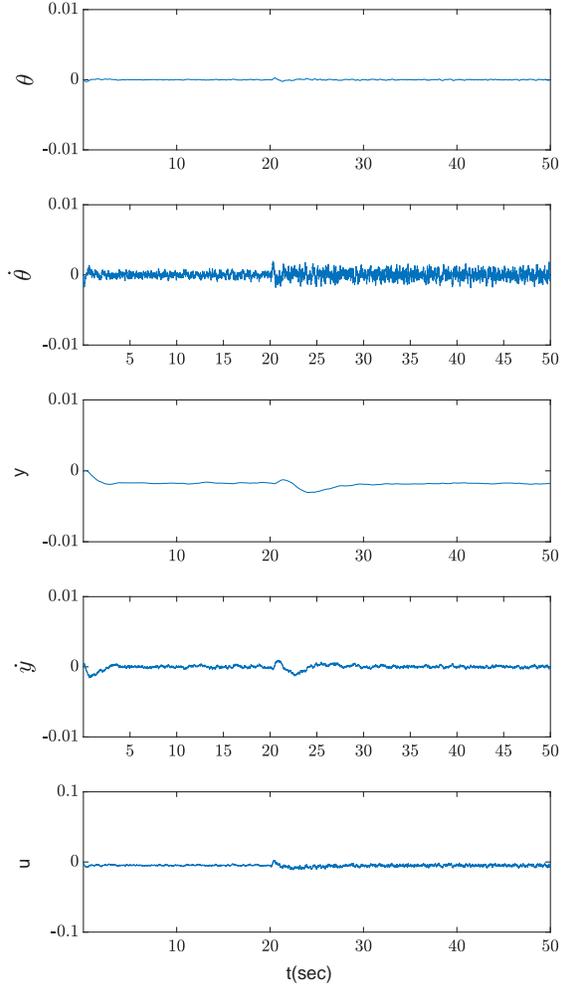

Figure 5 Time evaluation of states under sudden changes in parameters

As demonstrated by numerical experiments presented above, the devised LQR-based Q-learning algorithm can learn optimal controller gains in a few numbers of epochs and in the time scale of seconds without requiring extensive hyperparameter tuning. In contrast, existing DNN-based RL methods for inverted pendulum control typically require hundreds of epochs to converge [14], [15]. Moreover, each epoch takes a much longer time to finish due to

## 6. CONCLUSIONS

This study introduces a Q-learning-based LQR approach for balancing control of an inverted pendulum system. The proposed algorithm can learn the Q-function and optimal LQR controller gains without relying on a mathematical model. Instead, the algorithm can obtain optimal gains in real-time by interacting with the system through applying control inputs. Moreover, it can quickly adapt to parametric changes, as evidenced by the experimental results. In comparison to existing alternatives in the literature, the devised method is much more computationally efficient and does not require a large number of experiments for hyperparameter tuning.

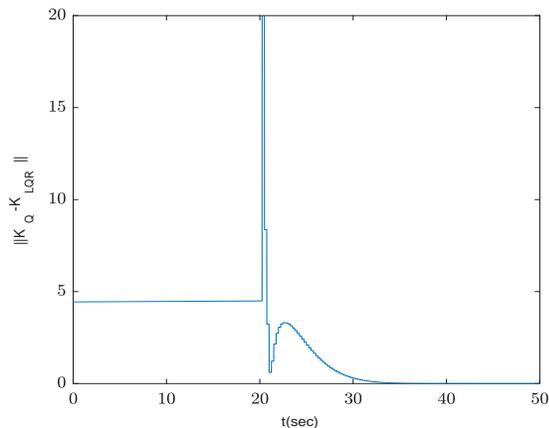

Figure 6 The norm of the difference between feedback gains adapted by the Q-learning-based LQR method ($K_Q$) under parameter changes and the gains of the model-based LQR method ($K_{LQR}$) obtained for new parameters